\documentclass[1,authoryear,12pt]{elsarticle}
\usepackage{lineno,hyperref}
\usepackage{graphicx}
\usepackage{subfigure}
\usepackage{amsmath,bm}
\usepackage{amsfonts}
\usepackage{amssymb}
\usepackage{amsbsy}
\usepackage{color}
\usepackage{multirow}
\usepackage{framed}
\usepackage{textcomp}
\usepackage{eurosym}
\usepackage{leftidx}
\usepackage{mathtools}
\usepackage{empheq}
\usepackage[top=30mm,left=30mm,body={155mm,210mm}]{geometry}
\usepackage{algorithm,algorithmic}
\usepackage[dvipsnames]{xcolor}
\usepackage[colorinlistoftodos, obeyFinal]{todonotes}
\usepackage{mdframed}
\usepackage{newfloat}
\usepackage{tabulary}

\makeatletter
\renewcommand{\ALG@name}{Box}
\makeatother
\DeclareFloatingEnvironment[
  fileext=lob,
  listname={List of Boxes},
  name=Box,
  placement=htp,
]{BOX}

\journal{}
\input{tcilatex}
\begin{document}

\begin{frontmatter}

\title{A multiplicative finite strain formulation for void growth models using elastic correctors}
\author[label1]{Meijuan Zhang}
\author[label2]{ Guadalupe Vadillo}
\author[label1]{Francisco Javier Montáns\corref{cor}}
\ead{fco.montans@upm.es}
\address[label1]{E.T.S.I Aeronautica y del Espacio, Pl. del Cardenal Cisneros, 3, 28040 Madrid, Spain}
\address[label2]{E.T.S.I Caminos Canales y Puertos, Cl. Profesor Aranguren, 3, 28040 Madrid, Spain}
\cortext[cor]{Corresponding author. }

\begin{abstract}

The proposed work is a formulation for large-strain non-isochoric plastic deformation, using the GTN yield function as the void growth rule as an example. This formulation is fully hyperelastic and uses the Kroner-Lee multiplicative decomposition. It adopts the concept of elastic correctors, and thus, does not have any constraints on the amount of elastic strain or the form of elasto-plastic behaviors. In addition, the volumetric part of the plastic deformation is described with the corrector of the volumetric part of the elastic logarithmic strain. It offers a new and sound kinematic approach to deal with non-isochoric plasticity. We also use the GTN function as an example to demonstrate the use of the kinematic relation as well as the implementation of the implicit algorithm. 
\end{abstract}


\end{frontmatter}

\section{\protect\bigskip Introduction}

Non-isochoric plastic deformation mechanisms like void growth are very
common. On the other hand, nowadays porous materials are widely used and
exist in different forms, for example, mechanical metamaterials are mostly
porous materials in the continuum scale. In addition, pores are the one of
the most common defects in material manufacture\citep{GAO2020}. The
prediction of void evolution can be very important for describing the
deformation and fracture mechanisms of such materials. One main approach is
by using the void growth models, which have two important aspects: one is
the void evolution rule, another one is the large-strain kinematics,
especially the relation between the volume change due to void evolution and
the strain type variable. The focus of this work is the latter.

The Gurson-Tvergaard-Needleman (GTN) type yield function is one of the most
widely used void evolution rule%
\citep{GURSON1977, TVERGAARD1984,
NEEDLEMAN1987}. There have been excessive extensions of the model, e.g., to
include the description of the nucleation and coalescence of the voids %
\citep{CHU1980,THOMSON2003,PARDOEN2000}, to account for the low or high
stress triaxiality case \citep{MALCHER2014,WU2019,XUE2008,VADILLO2016}, to
determine parameters \citep{ZHANG2021a,HE2021,YUE2022}, to avoid
mesh-dependance by non-local models \citep{HUTTER2013,BERGO2021}, to include
anisotropic yield criteria\citep{CHEN2009}, to consider cyclic loading
conditions\citep{WU2024}, and so on, however, there has not been much study
on the framework, especially there has not been an over-all sound and simple
framework for the large strain implementation. Most of the implementations
still use the small-strain formulations%
\citep{THOMSON2003,MALCHER2014,WU2019,VADILLO2016,ZHANG2021a,HE2021,WU2024},
which is not always appropriate, especially when void growth causes
localization of deformation which can easily exceed the small-strain regime.
Not only for the GTN models, no matter which void evolution rule is used,
the large-strain framework is fundamental for predicting accurately and
efficiently the void evolution.

Similar to the large-strain formulations for isochoric plasticity, in the
previous works, the large-strain kinematics of void growth models usually
lacks accuracy and generality, and the following are some examples. Hutter
et al. used a hypoelastic approach to relate the Jaumann stress rate and the
deformation rate \citep{HUTTER2013}, similar works can be found in %
\citep{NASIR2021,TUHAMI2022,MANSOURI2014,SEUPEL2020,ORSINI2001}, even though
hypoelastic formulations are known to be problematic \citep{SIMO1984}. Chen
et al. used the plastic logarithmic stains and its work conjugate to
calculate the dissipation \citep{CHEN2022}, which implementation is also not
overall sound, because instead of the Kroner-Lee multiplicative
decomposition, the authors simply used an additive decomposition of total
material logarithmic strain to get the elastic and plastic strain, a similar
treatment can be found in \citep{ZHANG2018}. Quinn et. al adopted the
conventional framework of crystal plasticity \citep{QUINN1997}, which
implementation is arguably complicated and involves several approximations
that result in constraints on the amount of elastic strain and elastoplastic
behaviors, and this approach can also be found in similar models that employ
crystal plasticity \citep{HA2010,POTIRNICHE2007,GUO2020,SHANG2020,LING2018}.
Mahnken in \citep{MAHNKEN1999} followed the work of Simo \citep{SIMO1992}
and used the formulation with Lie derivative of the elastic left
Gauchy-Green tensor, which is only valid for isotropic material and arguably
lacks simplicity. Bergo et al. used the Cauchy stress and the plastic
deformation rate to calculate the plastic dissipation \citep{BERGO2021},
which is also problematic because the work conjugate of the deformation rate
is the Kirchhoff stress.

Recently, we proposed a fully hyperelastic formulation for isochoric plastic
deformation that use elastic correctors 
\citep{LATORRE2018, ZHANG2019,
ZHANG2021b}, it maintains the Kroner-Lee multiplicative decomposition and
uses the logarithmic stain for achieving an additive structure. The
framework does not have any constraint on the amount of the elastic strain
and the form of elasto-plastic behaviors and it allows a simple
implementation of a fully hyperelastic formulation. In addition, it is
consistent and parallel to continuum elastoplasticity and crystal
plasticity. In this work, we extend our framework to void growth models,
moreover, we propose a new kinematic relation that relates the void
evolution with the elastic logarithmic strain. For simplicity, we use the
classic GTN\ yield function as the void evolution rule. Please note that
other evolution rules can also be applied. In the following part, we first
introduce the kinematic relations; and then the implementation of the GTN
model within this framework; we further demonstrate the soundness and
robustness of the implemented model by several numerical examples.

\section{Void growth kinematic relations}

The material consists of a solid matrix with volume $V_{m}$ which we
consider identical to the reference volume $V$, and a volume of voids of $%
v_{o}$. The current, locally unloaded volume is $V+v_{o}\equiv V_{m}+v_{o}$,
whereas an elastically deformed incremental volume $v_{e}$ gives a deformed
volume of the continuum equal to%
\begin{equation}
v=V_{m}+v_{o}+v_{e}
\end{equation}%
so we can write the functional dependence $v_{e}=v_{e}\left( v,v_{o}\right)
=\left( v-V_{m}\right) -v_{o}$, which in rate form is $\dot{v}_{e}=\dot{v}-%
\dot{v}_{o}$. Since pure plastic flow in the matrix is considered isochoric,
the change of volume $v_{p}$ due to plastic flow is assumed to come only
from the voids increment, i.e. $v_{p}=v_{o}$. Consider the deformation
gradient $\mathbf{X}$. The change of volume is given by the Jacobian of the
deformation, i.e. $J=\det \left( \mathbf{X}\right) $ which is

\begin{equation}
J=\frac{v}{V}=\frac{V_{m}+v_{o}+v_{e}}{V_{m}}=\left( \frac{V_{m}+v_{o}+v_{e}%
}{V_{m}+v_{o}}\right) \left( \frac{V_{m}+v_{o}}{V_{m}}\right)
=J_{e}J_{o}\equiv J_{e}J_{p}
\end{equation}%
where note that we obtain the multiplicative decomposition $%
J=J_{e}J_{o}\equiv J_{e}J_{p}$, which is the volumetric counterpart of%
\begin{equation}
\mathbf{X\equiv X}_{e}\underset{%
\begin{array}{c}
=:\mathbf{X}_{p}%
\end{array}%
}{\underbrace{\mathbf{X}_{o}\mathbf{X}_{p}^{d}}}=\mathbf{X}_{e}\mathbf{X}%
_{p}\Rightarrow J=\det \left( \mathbf{X}\right) =\det \left( \mathbf{X}%
_{e}\right) \det \left( \mathbf{X}_{p}\right) =J_{e}J_{p}
\end{equation}

\begin{figure}[tbp]
\centering
\includegraphics[width=0.7\textwidth]{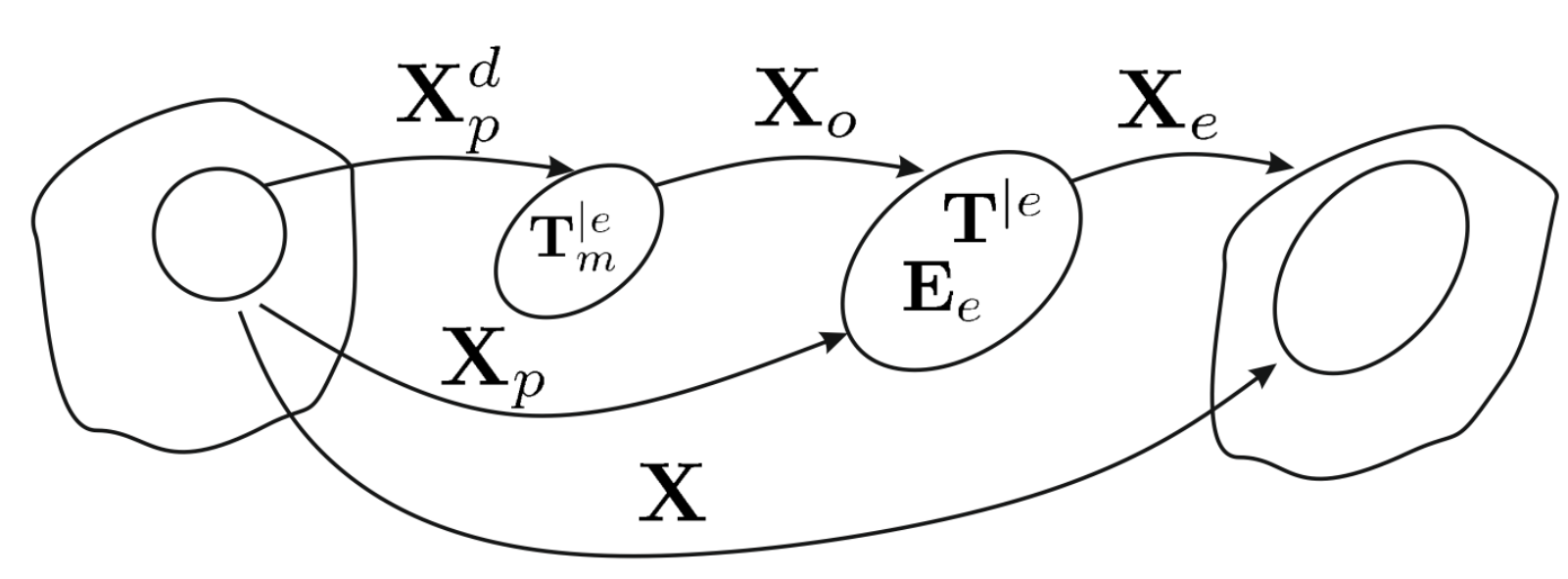} \label%
{multi_decomp}
\end{figure}

Even though they have some different nature, note that we incorporate $%
\mathbf{X}_{o}=J_{o}^{1/3}\mathbf{I}$ corresponding to the void growth to $%
\mathbf{X}_{p}^{d}$, corresponding to the (isochoric) plastic deformation of
the matrix. We define the void ratio as $f\left( v_{o}\right) =v_{o}/\left(
V_{m}+v_{o}\right) $. Then, using $J_{o}\equiv J_{p}=\left(
V_{m}+v_{o}\right) /V_{m}$ 
\begin{equation}
1-\frac{1}{J_{p}}=\frac{J_{p}-1}{J_{p}}=\frac{V_{m}+v_{o}}{V_{m}+v_{o}}-%
\frac{V_{m}}{V_{m}+v_{o}}=\frac{v_{o}}{V_{m}+v_{o}}=f
\end{equation}%
and%
\begin{equation}
1-f=\frac{V_{m}+v_{o}}{V_{m}+v_{o}}-\frac{v_{o}}{V_{m}+v_{o}}=\frac{V_{m}}{%
V_{m}+v_{o}}=J_{o}^{-1}  \label{1-f}
\end{equation}%
and%
\begin{equation}
\dot{f}=\frac{df}{dv_{o}}=\frac{\dot{v}_{o}}{V_{m}+v_{o}}-\frac{v_{o}\dot{v}%
_{o}}{\left( V_{m}+v_{o}\right) ^{2}}=\frac{V_{m}\dot{v}_{o}}{\left(
V_{m}+v_{o}\right) ^{2}}
\end{equation}%
so we obtain the following expression useful below%
\begin{equation}
\frac{\dot{f}}{1-f}=\frac{\dot{v}_{o}}{V_{m}+v_{o}}=\frac{\dot{J}_{p}}{J_{p}}
\end{equation}%
The Jacobian of the elastic deformation is

\begin{equation}
J_{e}=\frac{V_{m}+v_{o}+v_{e}}{V_{m}+v_{o}}
\end{equation}%
\begin{equation}
\dot{J}_{e}J_{e}^{-1}=\frac{\dot{v}_{e}}{V_{m}+v_{o}}+\frac{\dot{v}_{o}}{%
V_{m}+v_{o}}=\frac{\dot{v}_{e}}{V_{m}+v_{o}+v_{e}}+\frac{\dot{v}_{o}}{%
V_{m}+v_{o}+v_{e}}
\end{equation}%
The trial elastic rate of deformation, i.e. when the plastic flow is frozen
is%
\begin{equation}
\frac{d}{dt}\left. \left( J\right) \right\vert _{\dot{J}_{p}=\dot{v}_{o}=0}=%
\frac{d}{dt}\left. \left( \frac{V_{m}+v_{o}+v_{e}}{V_{m}}\right) \right\vert
_{\dot{J}_{p}=\dot{v}_{o}=0}=\frac{\dot{v}_{e}}{V_{m}}
\end{equation}%
so%
\begin{equation}
\frac{d}{dt}\left. \left( J\right) \right\vert _{\dot{J}_{p}=\dot{v}%
_{o}=0}J^{-1}=\frac{\dot{v}_{e}}{V_{m}}\frac{V_{m}}{V_{m}+v_{o}+v_{e}}=\frac{%
\dot{v}_{e}}{V_{m}+v_{o}+v_{e}}
\end{equation}%
Note that%
\begin{equation}
\left. \dot{J}_{e}\right\vert _{\dot{v}_{o}=0}J_{e}^{-1}=\frac{\dot{v}_{e}}{%
V_{m}+v_{o}}\frac{V_{m}+v_{o}}{V_{m}+v_{o}+v_{e}}=\frac{\dot{v}_{e}}{%
V_{m}+v_{o}+v_{e}}=\frac{d}{dt}\left. \left( J\right) \right\vert _{\dot{J}%
_{p}=\dot{v}_{o}=0}J^{-1}
\end{equation}%
hence, we obtain the relation consistent with the definition of elastic
trial rate%
\begin{equation}
\frac{\left. \dot{J}_{e}\right\vert _{\dot{v}_{o}=\dot{J}_{p}=0}}{J_{e}}=%
\frac{\left. \dot{J}\right\vert _{\dot{v}_{o}=\dot{J}_{p}=0}}{J}%
\Leftrightarrow ~^{tr}\dot{E}_{e}^{v}=\left. \dot{E}^{v}\right\vert _{\dot{v}%
_{o}=0}
\end{equation}%
Therefore, considering the case of only internal evolution, so $\dot{v}=0$,
we have%
\begin{equation}
\left. \frac{d}{dt}\left( J_{e}J_{p}\right) \right\vert _{\dot{v}=0}=\dot{J}%
_{e}J_{p}+J_{e}\dot{J}_{p}=0
\end{equation}%
we can also write%
\begin{equation}
\left[ \frac{d}{dt}\left( J_{e}J_{p}\right) \right] _{\dot{v}=0}\left(
J_{e}J_{p}\right) ^{-1}=0=\left. \frac{\dot{J}_{e}J_{p}+J_{e}\dot{J}_{p}}{%
J_{e}J_{p}}\right\vert _{\dot{v}=0}=\left. \frac{\dot{J}_{e}}{J_{e}}%
\right\vert _{\dot{v}=0}+\frac{\dot{J}_{p}}{J_{p}}
\end{equation}%
which means that, by definition of corrector rate ---note that $\dot{J}%
_{p}J_{p}^{-1}$ does not depend on $v$, but only on $v_{o}$%
\begin{equation}
~^{ct}\dot{E}_{e}^{v}=\left. \frac{\dot{J}_{e}}{J_{e}}\right\vert _{\dot{v}%
=0}=-\frac{\dot{J}_{p}}{J_{p}}
\end{equation}%
Hence%
\begin{equation}
\frac{\dot{J}_{e}}{J_{e}}=\dot{E}_{e}^{v}=\left. \dot{E}_{e}^{v}\right\vert
_{\dot{v}_{o}=0}+\left. \dot{E}_{e}^{v}\right\vert _{\dot{v}=0}=~^{tr}\dot{E}%
_{e}^{v}+~^{ct}\dot{E}_{e}^{v}=\left. \frac{\dot{J}}{J}\right\vert _{\dot{v}%
_{o}=\dot{J}_{p}=0}-\frac{\dot{J}_{p}}{J_{p}}
\end{equation}%
Then we obtain the following relation, \emph{of purely kinematic nature,
obtained by a simple relation of volumes,}%
\begin{equation}
-~^{ct}\dot{E}_{e}^{v}=\frac{\dot{J}_{p}}{J_{p}}\equiv \frac{\dot{J}_{o}}{%
J_{o}}=\frac{\dot{v}_{o}}{V_{m}+v_{o}}=\frac{\dot{f}}{1-f}
\label{vol of ct of Ee}
\end{equation}%
so we obtain a similar relation as in small strains, namely $tr\left( 
\mathbf{\dot{\varepsilon}}_{p}\right) =\dot{f}/\left( 1-f\right) $, c.f.
Eqs.(6) of \citep{GHOLIPOUR2019}, see also %
\citep{MALCHER2013,MALCHER2014,BERGO2021,ZHANG2021a}, but with the
interpretation of logarithmic elastic corrector rate, and with the use of
Jacobians. This is an accurate function, not got by using the mass
conservation principle in an approximate way by neglecting the volume change
due to elastic deformation like in small strains. Furthermore, note that
this is \emph{not} a constitutive relation, but a kinematic one \emph{that
must always hold}. Then, $\dot{f}$ \emph{is due to void change of any nature}
(i.e. including due to nucleation and shear). In fact, if we consider
different contributions ($\dot{f}_{g}$ is due to void growth, and $\dot{f}%
_{n}$ and $\dot{f}_{s}$ are due to nucleation and shear), $\dot{f}=\dot{f}%
_{g}+\dot{f}_{n}+\dot{f}_{s}$, we get%
\begin{equation}
-~^{ct}\dot{E}_{e}^{v}=\frac{\dot{f}_{g}+\dot{f}_{n}+\dot{f}_{s}}{1-f}=\frac{%
\dot{f}_{g}}{1-f}+\frac{\dot{f}_{n}}{1-f}+\frac{\dot{f}_{s}}{1-f}=-~^{ct,g}%
\dot{E}_{e}^{v}-~^{ct,n}\dot{E}_{e}^{v}-~^{ct,s}\dot{E}_{e}^{v}
\label{Ev decomp}
\end{equation}%
which consists of a corrector contribution of elastic volumetric strains due
to growth $~^{ct,g}\dot{E}_{e}^{v}$, one due to nucleation of new voids $%
~^{ct,n}\dot{E}_{e}^{v}$ and one due to shear interaction effects $~^{ct,s}%
\dot{E}_{e}^{v}$.

Consider now the change of volume of a void of volume $v_{o}$ and radius $r$%
. Then, $v_{o}=\frac{4}{3}\pi r^{3}$ and $\dot{v}_{o}/v_{o}=3\dot{r}/r$. Then%
\begin{equation}
\frac{\dot{v}_{o}}{v_{o}}\frac{v_{o}}{V+v_{o}}=3\frac{\dot{r}}{r}f=\dot{J}%
_{p}J_{p}^{-1}=-~^{ct}\dot{E}_{e}^{v}  \label{rdot/r with Ep}
\end{equation}

\bigskip The equation (\ref{vol of ct of Ee}) stands regardless of the
constitutive rule used. For example, when the void growth rule defined by
Rice and Tracey is used, neglecting for the moment the influence of the Lode
angle, the following expression for the growth rate of the void is given as
a function of the pressure $p$ and the yield stress in the matrix $\kappa $%
\begin{equation}
\frac{\dot{r}}{r}=\dot{\gamma}a\sinh \left( \frac{3}{2}\frac{p}{\kappa }%
\right)  \label{Rice Tracey}
\end{equation}%
where $a$ is a parameters ($\approx 0.56$), and $\dot{\gamma}$ is the rate
of effective strain (uniaxial equivalent). Since we interpret $\kappa $ as a
Kirchhoff-like yield stress in the intermediate configuration, $p$ is a
Kirchhoff-like pressure ($p/J$ is the Cauchy pressure, but note that for the
ratio $p/\kappa $ it is irrelevant if both stresses are of the same type).
Hence, using Eqs. (\ref{rdot/r with Ep}) and (\ref{Rice Tracey}) 
\begin{equation}
-~^{ct,g}\dot{E}_{e}^{v}=3\dot{\gamma}af\sinh \left( \frac{3}{2}\frac{p}{%
\kappa }\right) \;\Rightarrow \;-~^{ct,g}\mathbf{\dot{E}}_{e}^{v}=\tfrac{1}{3%
}~^{ct,g}\dot{E}_{e}^{v}\mathbf{I}=\dot{\gamma}af\sinh \left( \frac{3}{2}%
\frac{p}{\kappa }\right) \mathbf{I}  \label{EvRT}
\end{equation}

\section{Stored energy and dissipation}

The power in the solid by a volume vector load $\mathbf{b}$ and surface
traction $\mathbf{t}$ is (\citep{LATORRE2016})%
\begin{equation}
P_{V}=\int_{v}\mathbf{b}\cdot \mathbf{\dot{u}}dv+\int_{s}\mathbf{t}\cdot 
\mathbf{\dot{u}}ds=\int_{V}\mathbf{T}:\mathbf{\dot{E}}dV
\end{equation}%
where $\mathbf{T}$ is the work-conjugate stresses for referential
logarithmic stresses in the most general case. Note that $V$ is here the
reference volume. Then, if $\Psi \left( \mathbf{E}_{e}\right) $ is the
stored energy \emph{per reference volume (i.e. the stored energy in the
matrix per initial matrix volume)}, function of the elastic referential
strains $\mathbf{E}_{e}=\frac{1}{2}\ln \left( \mathbf{X}_{e}^{T}\mathbf{X}%
_{e}\right) $, then the dissipation \emph{per reference volume} $V$ is%
\begin{eqnarray}
\mathcal{\dot{D}}^{p} &=&\mathcal{P}-\dot{\Psi}=\mathbf{T}:\mathbf{\dot{E}}-%
\frac{d\Psi }{d\mathbf{E}_{e}}:\mathbf{\dot{E}}_{e}  \notag \\
&=&\mathbf{T}:\mathbf{\dot{E}}-\frac{d\Psi }{d\mathbf{E}_{e}}:\left( ~^{tr}%
\mathbf{\dot{E}}_{e}+~^{ct}\mathbf{\dot{E}}_{e}\right)
\end{eqnarray}%
As usual, we consider the two possible cases, which correspond to the two
partial derivatives considered, as a function of the two variables; recall $%
\mathbf{E}_{e}\left( \mathbf{E,X}_{p}\right) $. The first one is the absence
of dissipation, i.e. $\mathbf{\dot{X}}_{p}=~^{ct}\mathbf{\dot{E}}_{e}=%
\mathbf{0}$ and $\mathbf{\dot{E}}_{e}\equiv $ $^{tr}\mathbf{\dot{E}}_{e}$
and $\mathcal{D}^{p}=0$, so%
\begin{equation}
\mathbf{T}=\frac{d\Psi }{d\mathbf{E}_{e}}:\left. \frac{\partial \mathbf{E}%
_{e}}{\partial \mathbf{E}}\right\vert _{\mathbf{\dot{X}}_{p}=\mathbf{0}}=%
\mathbf{T}^{|e}:\left. \frac{\partial \mathbf{E}_{e}}{\partial \mathbf{E}}%
\right\vert _{\mathbf{\dot{X}}_{p}=\mathbf{0}}
\end{equation}%
where we have defined the internal stresses 
\begin{equation}
\mathbf{T}^{|e}=\frac{d\Psi }{d\mathbf{E}_{e}}
\end{equation}%
which are defined in the \emph{intermediate (\textquotedblleft
unloaded\textquotedblright ) configuration}\textbf{,} which is common to
that of the \textquotedblleft elastic\textquotedblright\ strains $\mathbf{E}%
^{e}$. Recall that the mapping $\left. \partial \mathbf{E}_{e}/\partial 
\mathbf{E}\right\vert _{\mathbf{\dot{X}}_{p}=\mathbf{0}}$ transforms the
pull-back to the reference configuration, where $\mathbf{T}$ lives.
Noteworthy, $\mathbf{T}$ for the case of isotropy and proportional loading
are the Kirchhoff stresses $\mathbf{\tau }=J\mathbf{\sigma }$, where $%
\mathbf{\sigma }$ are the Cauchy stresses.

As mentioned, the plastic deformation in the matrix is isochoric. There are
two types of irreversible deformations. One is due to the isochoric plastic
deformation of the matrix, given by $\mathbf{X}_{p}^{d}$, and one by void
growth given by $\mathbf{X}_{o}=J_{o}^{1/3}\mathbf{I}$, which is in essence
also a result of plastic flow in the matrix, but which manifests at the
continuum level through volumetric deformations. It can be also interpreted
as two modes of plastic deformation, one isochoric and the other one
volumetric. The decomposition of the deformation gradient is%
\begin{equation}
\mathbf{X}=\mathbf{X}_{e}\mathbf{X}_{o}\mathbf{X}_{p}^{d}\;\;\;\;\;\;\left( =%
\mathbf{X}_{e}\mathbf{X}_{p}\right)
\end{equation}%
This decomposition defines two intermediate configurations, see Figure \ref%
{multi_decomp}. The stress tensor $\mathbf{T}^{|e}$ lives in the
\textquotedblleft elastically unloaded\textquotedblright\ configuration.
However, that configuration includes the void growth.\emph{\ If a von Mises
yield criterion is applied to the matrix, the stresses should not be }$%
T^{|e} $\emph{, because the associated volume of reference is }$V+v_{o}$%
\emph{. In other words, the voids do not store energy. Then, using Eq. (\ref%
{1-f})}%
\begin{equation}
\mathbf{T}_{m}^{|e}:=\frac{v_{o}+V_{m}}{V_{m}}\mathbf{T}^{|e}=J_{o}\mathbf{T}%
^{|e}=\frac{\mathbf{T}^{|e}}{\left( 1-f\right) }
\end{equation}

After Tvergaard \citep{TVERGAARD1981} it is customary to include a sort of
stress concentration factor $q_{1}$ which amplifies the effect of the void
ratio, i.e. the effective (or fictitious continuum-equivalent) stress in the
matrix for plastic flow is%
\begin{equation}
\mathbf{T}_{m}^{|e}=\frac{\mathbf{T}^{|e}}{\left( 1-q_{1}f\right) }
\end{equation}%
where a typical value given is $q_{1}=1.5$, or $q_{1}=1.25$ accounting for
coalescence of voids. The dissipation in the matrix as given by the
equivalent stress is%
\begin{equation}
\mathcal{\dot{D}}_{m}^{p}=-\mathbf{T}_{m}^{|e}\mathbf{:}~^{ct}\mathbf{\dot{E}%
}_{e}\equiv -\mathbf{T}_{m}^{d|e}\mathbf{:}~^{ct}\mathbf{\dot{E}}%
_{e}^{d}=:\kappa \dot{\gamma}>0
\end{equation}%
where we considered $~^{ct}\mathbf{\dot{E}}_{e}$ be formed of a deviatoric
and a volumetric part $~^{ct}\mathbf{\dot{E}}_{e}=~^{ct}\mathbf{\dot{E}}%
_{e}^{d}+~^{ct}\mathbf{\dot{E}}_{e}^{v}$ and so $\mathbf{T}^{|e}=\mathbf{T}%
^{d|e}+\mathbf{T}^{v|e}$. 
\begin{equation}
~^{ct}\mathbf{\dot{E}}_{e}=~^{ct}\mathbf{\dot{E}}_{e}^{d}+~^{ct}\mathbf{\dot{%
E}}_{e}^{v}
\end{equation}%
The parameter $\kappa $ is the yield stress of the \emph{matrix} material
without voids, and $\dot{\gamma}$ is an equivalent plastic strain in the
matrix. Note that we can consider a constant yield stress $\kappa $ or
isotropic hardening given by $\kappa \left( \gamma \right) $ or by $\kappa
\left( D^{p}\right) $. We can write in this case the (associative) flow rule
as

\begin{equation}
-~^{ct}\mathbf{\dot{E}}_{e}^{d}=\sqrt{\frac{3}{2}}\dot{\gamma}\frac{\mathbf{T%
}_{m}^{d|e}}{\kappa }
\end{equation}%
so%
\begin{equation}
\dot{D}_{m}^{p}-\kappa \dot{\gamma}=\left( \frac{1}{\kappa }\sqrt{\frac{3}{2}%
}\mathbf{T}_{m}^{d|e}\mathbf{:T}_{m}^{d|e}-\kappa \right) \dot{\gamma}=0
\end{equation}%
or, if we use the stresses at the continuum level, taking into account some
porosity%
\begin{equation}
\dot{D}_{m}^{p}-\kappa \dot{\gamma}=\left( \frac{1}{\kappa }\sqrt{\frac{3}{2}%
}\frac{\mathbf{T}^{d|e}}{\left( 1-q_{1}f\right) }\mathbf{:}\frac{\mathbf{T}%
^{d|e}}{\left( 1-q_{1}f\right) }-\kappa \right) \dot{\gamma}=0
\end{equation}%
which, if $\dot{\gamma}>0$, it is equivalent to%
\begin{equation}
F^{d}=\frac{\frac{3}{2}\left\Vert \mathbf{T}^{d|e}\right\Vert ^{2}}{\kappa
^{2}}-\left( 1-q_{1}f\right) ^{2}=0
\end{equation}%
Note that this is the expression of the Gurson yield function given by
Tvergaard \&\ Needleman \citep{TVERGAARD1984} when there is no pressure
influence, i.e. when the pressure does not increase the void fraction. It is
often recognized that this function represents Lemaitre's damage model, from
continuum damage mechanics, in which $D:=q_{1}f$ is the Rabotnov damage
variable.

\section{Example: the GTN model}

In this work, we will focus on elaborating the use of the framework with the
classic and widely used GTN yield function as the constitutive rule which
has been extensively studied and used in modelling void growth. The previous
models are commonly implemented with small-strain framework. GTN function is
a yield function that relates the void growth rate with the isochoric
plasticity deformation.

\begin{equation}
F=\frac{\frac{3}{2}\left\Vert \mathbf{T}^{d|e}\right\Vert ^{2}}{\kappa ^{2}}%
-\left( 1-q_{1}f\right) ^{2}+2q_{1}f\left[ \cosh \left( q_{2}\frac{3}{2}%
\frac{p}{\kappa }\right) -1\right] =0  \label{GTN function}
\end{equation}%
where $q_{1}$ is a parameter that amplifies the effect of void ratio, and $%
q_{2}$ is another parameter frequently set to $q_{2}=1$, and for the
pressure of the continuum $p=0$ the term in brackets vanishes. The parameter 
$\kappa $ is the yield stress of the \emph{matrix} material without voids. $%
\mathbf{T}^{d|e}$ is de deviatoric stress of the continuum in the
intermediate configuration. Noteworthy, the derivatives respect to the
pressure and to $\mathbf{T}^{d|e}$ are 
\begin{equation}
\frac{\partial F}{\partial p}=\frac{3q_{2}\left( q_{1}f\right) }{\kappa }%
\sinh \left( \frac{3}{2}\frac{p}{\kappa }q_{2}\right) \text{ \ and \ }\frac{%
\partial F}{\partial \mathbf{T}^{d|e}}=3\frac{\mathbf{T}^{d|e}}{\kappa ^{2}}
\end{equation}%
Then, using the assumed associativity of \textquotedblleft flow
rule\textquotedblright\ for the continuum, 
\begin{eqnarray}
-~^{ct,g}\mathbf{\dot{E}}_{e} &=&\dot{\lambda}\frac{dF}{d\mathbf{T}^{|e}}=%
\dot{\lambda}\frac{dF}{d\mathbf{T}^{d|e}}:\frac{d\mathbf{T}^{d|e}}{d\mathbf{T%
}^{|e}}+\dot{\lambda}\frac{dF}{dp}\frac{dp}{d\mathbf{T}^{|e}}
\label{flow rule total} \\
&=&3\dot{\lambda}\frac{\mathbf{T}^{d|e}}{\kappa ^{2}}+\dot{\lambda}\frac{%
q_{2}\left( q_{1}f\right) }{\kappa }\sinh \left( \frac{3}{2}\frac{p}{\kappa }%
q_{2}\right) \mathbf{I}  \notag \\
&=&-~^{ct}\mathbf{\dot{E}}_{e}^{d}-~^{ct,g}\mathbf{\dot{E}}_{e}^{v}=-~^{ct}%
\mathbf{\dot{E}}_{e}^{d}-\frac{1}{3}~^{ct,g}\dot{E}_{e}^{v}\mathbf{I}
\end{eqnarray}

\bigskip Here $\lambda $ is the equivalent plastic strain in the continuum.
Note that we can consider a constant yield stress $\kappa $ or isotropic
hardening given by $\kappa \left( \lambda \right) $ or by the accumulated
plastic strain. In this work, we add isotropic hardening by using a simple
linear hardening function $\kappa (\lambda )=\kappa _{0}+H\lambda $, with $%
\kappa _{0}$ as the yield stress without hardening, and $H$ as the hardening
parameter.

The deviatoric and volumetric part of the corrector of elastic strain rate
are:

\begin{equation}
\mathbf{-}^{ct}\mathbf{\dot{E}}_{e}^{d}=3\dot{\lambda}\frac{\mathbf{T}%
^{d\mid e}}{\kappa ^{2}}  \label{Eect_dev}
\end{equation}

\begin{equation}
\mathbf{-}^{ct}\mathbf{\dot{E}}_{e}^{v}=\frac{q_{1}q_{2}\dot{\lambda}f}{%
\kappa }\sinh (\frac{3}{2}\frac{p}{\kappa }q_{2})\mathbf{I}  \label{Eect_vol}
\end{equation}

Considering that from the kinematics, we already get (\ref{vol of ct of Ee}%
), so when $^{ct}\dot{E}_{e}^{v}\neq 0$, we have

\begin{equation}
\mathbf{-}^{ct}\mathbf{\dot{E}}_{e}^{v}=\frac{q_{1}q_{2}\dot{\lambda}f}{%
\kappa }\sinh (\frac{3}{2}\frac{p}{\kappa }q_{2})\mathbf{I}=\frac{1}{3}\frac{%
\dot{f}}{1-f}\mathbf{I}
\end{equation}

\bigskip It is noteworthy that when using current GTN yield function, when
there is no volume change, there is no void evolution, so when only under
isochoric deformation, there is no volume change of voids. A lot of improved
GTN models have been proposed to account for the low stress triaxiality case %
\citep{MALCHER2014,XUE2008,WU2019}, but it is not the purpose of this work
to extend a GTN model to account for specific cases.

\subsection{The algorithm for GTN model}

The main calculation is carried out in the intermediate configuration. The
integration algorithm is similar to that of small-strain formulation, only a
post-processor would be added to change the stress and the elastoplastic
tangent to the desired measure in the desired configuration. In the
following part we are going to introduce the integration algorithm including
isotropic hardening.

The trial of the logarithmic strain in the intermediate configuration of
current step $^{t+\Delta t}\mathbf{E}^{\mid e}$ can be obtained from the
plastic deformation gradient of last step $^{t}\mathbf{X}_{p}$ and the
deformation gradient of current step $^{t+\Delta t}\mathbf{X}$:

\begin{equation}
^{tr}\mathbf{X}_{e}=^{t+\Delta t}\mathbf{X}^{t}\mathbf{X}_{p}^{-1}\text{ \ \
\ \ \ \ \ \ \ \ \ \ \ \ }^{tr}\mathbf{E}^{\mid e}=\frac{1}{2}\ln (^{tr}%
\mathbf{X}_{e}^{Ttr}\mathbf{X}_{e})\text{\ }
\end{equation}

To calculate the corrector $^{ct}\mathbf{E}^{\mid e}$, there are several
alternative ways to carry out the local iterations, here we choose $%
^{ct}E_{e}^{v}$ and $^{ct}\mathbf{E}_{e}^{d}$ as the variables for the
iterations. From the yield function (\ref{GTN function}) and the flow rule (%
\ref{flow rule total}), we get the deviatoric (\ref{Eect_dev}) and
volumetric part (\ref{Eect_vol}) of the corrector of the elastic strain
rate, and if we represent $\Delta \lambda $ with $^{ct}E_{e}^{v}$:

\begin{equation}
\Delta \lambda =-\frac{^{t+\Delta t}\kappa ^{ct}E_{e}^{v}}{%
3q_{1}q_{2}^{t+\Delta t}f\sinh (\frac{3q_{2}}{2}\frac{^{t+\Delta t}p}{%
^{t+\Delta t}\kappa })}  \label{dlam_with_Eectvol}
\end{equation}

By substituting the kinematic relation \ref{vol of ct of Ee}, we can get:

\begin{equation}
\Delta \lambda =\frac{^{t+\Delta t}\kappa \Delta f}{3q_{1}q_{2}^{t+\Delta
t}f(1-^{t+\Delta t}f)\sinh (\frac{3q_{2}}{2}\frac{^{t+\Delta t}p}{^{t+\Delta
t}\kappa })}
\end{equation}

After substituting it into (\ref{Eect_dev}), we get an residual function $G$
for $^{ct}\mathbf{E}_{e}^{d}$. So for the local iterations, the two residual
functions for the Newton-Raphson method can be:

\begin{equation}
^{t+\Delta t}F=\frac{\frac{3}{2}\left\Vert ^{t+\Delta t}\mathbf{T}^{d\mid
e}\right\Vert ^{2}}{^{t+\Delta t}\kappa ^{2}}-(1-q_{1}^{t+\Delta
t}f)^{2}+2q_{1}^{t+\Delta t}f\text{ }[\cosh (\frac{3q_{2}}{2}\frac{%
^{t+\Delta t}p}{^{t+\Delta t}\kappa })-1]=0
\end{equation}

and

\begin{equation}
^{t+\Delta t}G:=^{ct}\mathbf{E}_{e}^{d}-\frac{^{ct}E_{e}^{v}}{^{t+\Delta
t}\kappa q_{1}q_{2}^{t+\Delta t}f\sinh (\frac{3q_{2}}{2}\frac{^{t+\Delta t}p%
}{^{t+\Delta t}\kappa })}^{t+\Delta t}\mathbf{T}^{d\mid e}=\mathbf{0}
\label{residual function G}
\end{equation}

Because $^{t+\Delta t}\mathbf{T}^{\mid e}-^{t+\Delta t}\mathbb{A}^{\mid
e}:(^{tr}\mathbf{E}^{\mid e}+^{ct}\mathbf{E}^{v\mid e}+^{ct}\mathbf{E}%
^{d\mid e})=\mathbf{0}$, and $^{t+\Delta t}\mathbf{T}^{\mid e}=^{t+\Delta t}%
\mathbf{T}^{d\mid e}+^{t+\Delta t}p\mathbf{I}$, we have:

\begin{equation}
^{t+\Delta t}\mathbf{T}^{d\mid e}=^{tr}\mathbf{T}^{d\mid e}+^{t+\Delta t}%
\mathbb{A}^{\mid e}:^{ct}\mathbf{E}^{d\mid e}
\end{equation}

\begin{equation}
^{t+\Delta t}p\mathbf{I}\mathbf{=}^{tr}\mathbf{T}^{v\mid e}+^{t+\Delta t}%
\mathbb{A}^{\mid e}:^{ct}\mathbf{E}^{v\mid e}
\end{equation}

Since we assume that isotropic hardening is linear, we have:

\begin{equation}
\frac{d^{t+\Delta t}\kappa }{d\Delta \lambda }=H  \label{dk_ddlam}
\end{equation}

\begin{eqnarray}
\frac{\partial ^{t+\Delta t}\kappa }{\partial ^{ct}E^{v\mid e}} &=&\frac{%
\partial ^{t+\Delta t}\kappa }{\partial \Delta \lambda }\frac{\partial
\Delta \lambda }{\partial ^{ct}E^{v\mid e}} \\
&=&\frac{\partial ^{t+\Delta t}\kappa }{\partial \Delta \lambda }(\frac{%
\partial \Delta \lambda }{\partial ^{t+\Delta t}f}\frac{\partial ^{t+\Delta
t}f}{\partial ^{ct}E^{v\mid e}}+\frac{\partial \Delta \lambda }{\partial
^{t+\Delta t}p}\frac{\partial ^{t+\Delta t}p}{\partial ^{ct}E^{v\mid e}}+%
\frac{\partial \Delta \lambda }{\partial ^{t+\Delta t}\kappa }\frac{\partial
^{t+\Delta t}\kappa }{\partial ^{ct}E^{v\mid e}})  \notag
\end{eqnarray}

After rearranging it, we can get $\frac{\partial ^{t+\Delta t}\kappa }{%
\partial ^{ct}E^{v\mid e}}$:

\begin{equation}
\frac{\partial ^{t+\Delta t}\kappa }{\partial ^{ct}E^{v\mid e}}=(1-\frac{%
d^{t+\Delta t}\kappa }{d\Delta \lambda }\frac{\partial \Delta \lambda }{%
\partial ^{t+\Delta t}\kappa })^{-1}(\frac{d^{t+\Delta t}\kappa }{d\Delta
\lambda }\frac{\partial \Delta \lambda }{\partial ^{t+\Delta t}f}\frac{%
\partial ^{t+\Delta t}f}{\partial ^{ct}E^{v\mid e}}+\frac{d^{t+\Delta
t}\kappa }{d\Delta \lambda }\frac{\partial \Delta \lambda }{\partial
^{t+\Delta t}p}\frac{\partial ^{t+\Delta t}p}{\partial ^{ct}E^{v\mid e}})
\end{equation}

With the derivatives:

\begin{align}
\frac{\partial \Delta \lambda }{\partial ^{t+\Delta t}\kappa }& =\frac{%
\Delta f}{3q_{1}q_{2}^{t+\Delta t}f(1-^{t+\Delta t}f)\sinh (\frac{3q_{2}}{2}%
\frac{^{t+\Delta t}p}{^{t+\Delta t}\kappa })}  \notag \\
& +\frac{\Delta f\cosh (\frac{3q_{2}}{2}\frac{^{t+\Delta t}p}{^{t+\Delta
t}\kappa })^{t+\Delta t}p}{2q_{1}^{t+\Delta t}\kappa ^{t+\Delta
t}f(1-^{t+\Delta t}f)\sinh ^{2}(\frac{3q_{2}}{2}\frac{^{t+\Delta t}p}{%
^{t+\Delta t}\kappa })}
\end{align}

\begin{align}
\frac{\partial \Delta \lambda }{\partial ^{t+\Delta t}f}&=\frac{^{t+\Delta
t}\kappa }{3q_{1}q_{2}^{t+\Delta t}f(1-^{t+\Delta t}f)\sinh (\frac{3q_{2}}{2}%
\frac{^{t+\Delta t}p}{^{t+\Delta t}\kappa })}  \notag \\
& - \frac{\Delta f^{t+\Delta t}\kappa (1-2^{t+\Delta t}f)}{%
3q_{1}q_{2}^{t+\Delta t}f^{2}(1-^{t+\Delta t}f)^{2}\sinh (\frac{3q_{2}}{2}%
\frac{^{t+\Delta t}p}{^{t+\Delta t}\kappa })}
\end{align}

\begin{equation}
\frac{\partial \Delta \lambda }{\partial ^{t+\Delta t}p}=-\frac{\Delta
f\cosh (\frac{3q_{2}}{2}\frac{^{t+\Delta t}p}{^{t+\Delta t}\kappa })}{%
2q_{1}^{t+\Delta t}f(1-^{t+\Delta t}f)\sinh ^{2}(\frac{3q_{2}}{2}\frac{%
^{t+\Delta t}p}{^{t+\Delta t}\kappa })}
\end{equation}

The derivatives of the first residual function $^{t+\Delta t}F$ are:

\begin{equation}
\frac{\partial ^{t+\Delta t}F}{\partial ^{ct}\mathbf{E}^{d\mid e}}=\frac{%
\partial ^{t+\Delta t}F}{\partial ^{t+\Delta t}\mathbf{T}^{d\mid e}}\frac{%
\partial ^{t+\Delta t}\mathbf{T}^{d\mid e}}{\partial ^{ct}\mathbf{E}^{d\mid
e}}
\end{equation}

\begin{align}
\frac{\partial ^{t+\Delta t}F}{\partial ^{ct}E^{v\mid e}}&=\frac{\partial
^{t+\Delta t}F}{\partial ^{t+\Delta t}p}\frac{\partial ^{t+\Delta t}p}{%
\partial ^{t+\Delta t}\mathbf{T}^{v\mid e}}\frac{\partial ^{t+\Delta t}%
\mathbf{T}^{v\mid e}}{\partial ^{ct}\mathbf{E}^{v\mid e}}\frac{\partial ^{ct}%
\mathbf{E}^{v\mid e}}{\partial ^{ct}E^{v\mid e}}  \notag \\
&+\frac{\partial ^{t+\Delta t}F}{\partial ^{t+\Delta t}f}\frac{\partial
^{t+\Delta t}f}{\partial ^{ct}E^{v\mid e}}+\frac{\partial ^{t+\Delta t}F}{%
\partial ^{t+\Delta t}\kappa }\frac{\partial ^{t+\Delta t}\kappa }{\partial
^{ct}E^{v\mid e}}
\end{align}

With the derivatives: 
\begin{equation}
\frac{\partial ^{t+\Delta t}F}{\partial ^{t+\Delta t}\mathbf{T}^{d\mid e}}=%
\frac{3^{t+\Delta t}\mathbf{T}^{d\mid e}}{^{t+\Delta t}\kappa ^{2}}
\end{equation}

\begin{equation}
\frac{\partial ^{t+\Delta t}\mathbf{T}^{d\mid e}}{\partial ^{ct}\mathbf{E}%
^{d\mid e}}=^{t+\Delta t}\mathbb{A}^{\mid e}
\end{equation}

\begin{equation}
\frac{\partial ^{t+\Delta t}F}{\partial ^{t+\Delta t}p}=\frac{%
3q_{1}q_{2}^{t+\Delta t}f}{^{t+\Delta t}\kappa }\sinh (\frac{3q_{2}}{2}\frac{%
^{t+\Delta t}p}{^{t+\Delta t}\kappa })
\end{equation}

\begin{equation}
\frac{\partial ^{t+\Delta t}p}{\partial ^{ct}E^{v\mid e}}=\frac{\partial
^{t+\Delta t}p}{\partial ^{t+\Delta t}\mathbf{T}^{v\mid e}}\frac{\partial
^{t+\Delta t}\mathbf{T}^{v\mid e}}{\partial ^{ct}\mathbf{E}^{v\mid e}}\frac{%
\partial ^{ct}\mathbf{E}^{v\mid e}}{\partial ^{ct}E^{v\mid e}}=\frac{1}{9}%
\mathbf{I:}^{t+\Delta t}\mathbb{A}^{\mid e}\mathbf{:I}
\end{equation}

\begin{equation}
\frac{\partial ^{t+\Delta t}F}{\partial ^{t+\Delta t}f}=2q_{1}\cosh (\frac{%
3q_{2}}{2}\frac{^{t+\Delta t}p}{^{t+\Delta t}\kappa })-2q_{1}^{t+\Delta t}f
\end{equation}

\begin{equation}
\frac{\partial ^{t+\Delta t}f}{\partial ^{ct}E^{v\mid e}}=1/[\frac{1}{%
^{t+\Delta t}f-1}-\frac{\Delta f}{(^{t+\Delta t}f-1)^{2}}]
\end{equation}

\begin{equation}
\frac{\partial ^{t+\Delta t}F}{\partial ^{t+\Delta t}\kappa }=-\frac{%
3\left\Vert ^{t+\Delta t}\mathbf{T}^{d\mid e}\right\Vert ^{2}}{^{t+\Delta
t}\kappa ^{3}}-\frac{3q_{1}q_{2}^{t+\Delta t}f^{t+\Delta t}p\sinh (\frac{%
3q_{2}}{2}\frac{^{t+\Delta t}p}{^{t+\Delta t}\kappa })}{^{t+\Delta t}\kappa
^{2}}
\end{equation}

The derivatives of the second residual function $^{t+\Delta t}G$ are:

\begin{equation}
\frac{\partial ^{t+\Delta t}G}{\partial ^{ct}\mathbf{E}^{d\mid e}}%
=[^{t+\Delta t}\kappa q_{1}q_{2}^{t+\Delta t}f\sinh (\frac{3q_{2}}{2}\frac{%
^{t+\Delta t}p}{^{t+\Delta t}\kappa })]\mathbb{I}+\frac{\partial ^{t+\Delta
t}G}{\partial ^{t+\Delta t}\mathbf{T}^{d\mid e}}\frac{\partial ^{t+\Delta t}%
\mathbf{T}^{d\mid e}}{\partial ^{ct}\mathbf{E}^{d\mid e}}
\end{equation}

\begin{equation}
\frac{\partial ^{t+\Delta t}G}{\partial ^{ct}E^{v\mid e}}=-^{t+\Delta t}%
\mathbf{T}^{d\mid e}+\frac{\partial ^{t+\Delta t}G}{\partial ^{t+\Delta t}p}%
\frac{\partial ^{t+\Delta t}p}{\partial ^{ct}E^{v\mid e}}+\frac{\partial
^{t+\Delta t}G}{\partial ^{t+\Delta t}f}\frac{\partial ^{t+\Delta t}f}{%
\partial ^{ct}E^{v\mid e}}+\frac{\partial ^{t+\Delta t}G}{\partial
^{t+\Delta t}\kappa }\frac{\partial ^{t+\Delta t}\kappa }{\partial
^{ct}E^{v\mid e}}
\end{equation}

With the derivatives:

\begin{equation}
\frac{\partial ^{t+\Delta t}G}{\partial ^{t+\Delta t}\mathbf{T}^{d\mid e}}%
=-^{ct}E_{e}^{v}
\end{equation}

\begin{equation}
\frac{\partial ^{t+\Delta t}G}{\partial ^{t+\Delta t}p}=\frac{3}{2}%
q_{1}q_{2}^{2t+\Delta t}f\cosh (\frac{3q_{2}}{2}\frac{^{t+\Delta t}p}{%
^{t+\Delta t}\kappa })^{ct}\mathbf{E}_{e}^{d}
\end{equation}

\begin{equation}
\frac{\partial ^{t+\Delta t}G}{\partial ^{t+\Delta t}f}=[^{t+\Delta t}\kappa
q_{1}q_{2}\sinh (\frac{3q_{2}}{2}\frac{^{t+\Delta t}p}{^{t+\Delta t}\kappa }%
)]^{ct}\mathbf{E}_{e}^{d}
\end{equation}

\begin{equation}
\frac{\partial ^{t+\Delta t}G}{\partial ^{t+\Delta t}\kappa }%
=q_{1}q_{2}^{t+\Delta t}f\sinh (\frac{3q_{2}}{2}\frac{^{t+\Delta t}p}{%
^{t+\Delta t}\kappa })^{ct}\mathbf{E}_{e}^{d}-\frac{3q_{1}q_{2}^{2t+\Delta
t}p^{t+\Delta t}f}{2^{t+\Delta t}\kappa }\cosh (\frac{3q_{2}}{2}\frac{%
^{t+\Delta t}p}{^{t+\Delta t}\kappa })^{ct}\mathbf{E}_{e}^{d}
\end{equation}

A plain Newton-Raphson method is used for the iteration, and the Jacobian
matrix is:

\begin{equation}
\begin{bmatrix}
\dfrac{\partial ^{t+\Delta t}F}{\partial ^{ct}\mathbf{E}^{d\mid e}} & \dfrac{%
\partial ^{t+\Delta t}F}{\partial ^{ct}E^{v\mid e}} \\ 
\dfrac{\partial ^{t+\Delta t}G}{\partial ^{ct}\mathbf{E}^{d\mid e}} & \dfrac{%
\partial ^{t+\Delta t}G}{\partial ^{ct}E^{v\mid e}}%
\end{bmatrix}%
\end{equation}

For updating the variables, an extra iteration is necessary when isotropic
hardening is included. The variables of the local iterations are $^{ct}%
\mathbf{E}^{d\mid e}$ and $^{ct}E^{v\mid e}$, and other variables obtained
directly are $\Delta f$ and $^{t+\Delta t}\mathbf{T}^{\mid e}$, but because $%
\Delta \lambda $ and $^{t+\Delta t}\kappa $ are dependant on each other, an
extra iteration is needed to update them.

To get the analytical tangent, the key is to get the derivative of $^{ct}%
\mathbf{E}_{e}^{d}$ and $^{ct}\mathbf{E}_{e}^{v}$ to $_{0}^{t+\Delta t}%
\mathbf{E}_{e}$. Because $_{0}^{t+\Delta t}\mathbf{E}_{e}=^{tr}\mathbf{E}%
_{e}+^{ct}\mathbf{E}_{e}$, the derivative to $^{tr}\mathbf{E}_{e}$ is:

\begin{equation}
\frac{d\,_{0}^{t+\Delta t}\mathbf{E}_{e}}{d\,^{tr}\mathbf{E}_{e}}=\mathbb{I+}%
\frac{d^{ct}\mathbf{E}_{e}}{d\,_{0}^{t+\Delta t}\mathbf{E}_{e}}\frac{%
d\,_{0}^{t+\Delta t}\mathbf{E}_{e}}{d\,^{tr}\mathbf{E}_{e}}
\end{equation}

\begin{equation}
\frac{d\,_{0}^{t+\Delta t}\mathbf{E}_{e}}{d\,^{tr}\mathbf{E}_{e}}=(\mathbb{I-%
}\frac{d^{ct}\mathbf{E}_{e}}{d\,_{0}^{t+\Delta t}\mathbf{E}_{e}})^{-1}
\label{dEe_dEetr}
\end{equation}

\begin{eqnarray}
\frac{d\,^{ct}\mathbf{E}_{e}}{d\,_{0}^{t+\Delta t}\mathbf{E}_{e}} &=&\frac{%
d\,^{ct}\mathbf{E}_{e}^{d}}{d\,_{0}^{t+\Delta t}\mathbf{E}_{e}}+\frac{%
d\,^{ct}\mathbf{E}_{e}^{v}}{d\,_{0}^{t+\Delta t}\mathbf{E}_{e}}
\label{dEect_dEe} \\
&=&\frac{d\,^{ct}\mathbf{E}_{e}^{d}}{d\,_{0}^{t+\Delta t}\mathbf{E}_{e}}+%
\frac{d\,^{ct}\mathbf{E}_{e}^{v}}{d^{t+\Delta t}f}\frac{d^{t+\Delta t}f}{%
d_{0}^{t+\Delta t}\mathbf{E}_{e}}  \notag
\end{eqnarray}

First, for the volumetric part, because of the kinematic relation of the
volumetric part \ref{vol of ct of Ee}, we obtain $\frac{d^{ct}\mathbf{E}%
^{v\mid e}}{d\,^{t+\Delta t}f}$ as:

\begin{equation}
\frac{d^{ct}\mathbf{E}^{v\mid e}}{d^{t+\Delta t}\,f}=\frac{1}{3}\mathbf{I(}%
\frac{1}{^{t+\Delta t}f-1}-\frac{\Delta f}{(^{t+\Delta t}f-1)^{2}}\mathbf{)}
\end{equation}

Second, for the deviatoric part, because $\mathbf{-}^{ct}\mathbf{\dot{E}}%
_{e}^{d}=3\dot{\lambda}\frac{\mathbf{T}^{d\mid e}}{\kappa ^{2}}$, the
derivative $\frac{d^{ct}\mathbf{E}^{d\mid e}}{d\,_{0}^{t+\Delta t}\mathbf{E}%
_{e}}$ can be written as:

\begin{align}
\frac{d^{ct}\mathbf{E}^{d\mid e}}{d\,_{0}^{t+\Delta t}\mathbf{E}_{e}}&=\frac{%
\partial ^{ct}\mathbf{E}^{d\mid e}}{\partial ^{t+\Delta t}\mathbf{T}^{d\mid
e}}\frac{d^{t+\Delta t}\mathbf{T}^{d\mid e}}{d^{t+\Delta t}\mathbf{T}^{\mid
e}}\frac{d^{t+\Delta t}\mathbf{T}^{\mid e}}{d\,_{0}^{t+\Delta t}\mathbf{E}%
_{e}}+\frac{\partial ^{ct}\mathbf{E}^{d\mid e}}{\partial \Delta \lambda }%
\frac{d\Delta \lambda }{d\,_{0}^{t+\Delta t}\mathbf{E}_{e}}  \notag \\
&+\frac{\partial ^{ct}\mathbf{E}^{d\mid e}}{\partial ^{t+\Delta t}\kappa }%
\frac{d^{t+\Delta t}\kappa }{d^{ct}E^{v\mid e}}\frac{d^{ct}E^{v\mid e}}{%
d^{t+\Delta t}f}\frac{d^{t+\Delta t}f}{d_{0}^{t+\Delta t}\mathbf{E}_{e}}
\label{dEectdev_dEe}
\end{align}

Here the key is to get $\frac{d\Delta \lambda }{d\,_{0}^{t+\Delta t}\mathbf{E%
}_{e}}$ and $\frac{d^{t+\Delta t}f}{d_{0}^{t+\Delta t}\mathbf{E}_{e}}$, they
can be calculated at the same time from the expression of $\Delta \lambda $
and the yield function. Because: 
\begin{equation}
\Delta \lambda =-\frac{^{t+\Delta t}\kappa ^{ct}E^{v\mid e}}{%
3q_{1}q_{2}^{t+\Delta t}f\sinh (\frac{3q_{2}}{2}\frac{^{t+\Delta t}p}{%
^{t+\Delta t}\kappa })}
\end{equation}

\begin{align}
\frac{d\Delta \lambda }{d\,_{0}^{t+\Delta t}\mathbf{E}_{e}}& =\frac{\partial
\Delta \lambda }{\partial ^{ct}E^{v\mid e}}\frac{d^{ct}E^{v\mid e}}{%
d^{t+\Delta t}f}\frac{d^{t+\Delta t}f}{d_{0}^{t+\Delta t}\mathbf{E}_{e}}+%
\frac{\partial \Delta \lambda }{\partial ^{t+\Delta t}f}\frac{d^{t+\Delta t}f%
}{d_{0}^{t+\Delta t}\mathbf{E}_{e}}  \notag \\
& +\frac{\partial \Delta \lambda }{\partial ^{t+\Delta t}\kappa }\frac{%
d^{t+\Delta t}\kappa }{d\Delta \lambda }\frac{d\Delta \lambda }{%
d_{0}^{t+\Delta t}\mathbf{E}_{e}}+\frac{\partial \Delta \lambda }{\partial
^{t+\Delta t}p}\frac{d^{t+\Delta t}p}{d^{t+\Delta t}\mathbf{T}^{\mid e}}%
\frac{d^{t+\Delta t}\mathbf{T}^{\mid e}}{d_{0}^{t+\Delta t}\mathbf{E}_{e}}
\label{(1)}
\end{align}

After rewriting (\ref{(1)}), we can get:

\begin{align}
(1-\frac{\partial \Delta \lambda }{\partial ^{t+\Delta t}\kappa }\frac{%
d^{t+\Delta t}\kappa }{d\Delta \lambda })\frac{d\Delta \lambda }{%
d\,_{0}^{t+\Delta t}\mathbf{E}_{e}}&=(\frac{\partial \Delta \lambda }{%
\partial ^{ct}E^{v\mid e}}\frac{d^{ct}E^{v\mid e}}{d^{t+\Delta t}f}+\frac{%
\partial \Delta \lambda }{\partial ^{t+\Delta t}f})\frac{d^{t+\Delta t}f}{%
d_{0}^{t+\Delta t}\mathbf{E}_{e}}  \notag \\
&+\frac{\partial \Delta \lambda }{\partial ^{t+\Delta t}p}\frac{d^{t+\Delta
t}p}{d^{t+\Delta t}\mathbf{T}^{\mid e}}\frac{d^{t+\Delta t}\mathbf{T}^{\mid
e}}{d_{0}^{t+\Delta t}\mathbf{E}_{e}}
\end{align}

The derivatives used are \ref{dk_ddlam} and:%
\begin{equation}
\frac{\partial \Delta \lambda }{\partial ^{ct}E^{v\mid e}}=-\frac{^{t+\Delta
t}\kappa }{3q_{1}q_{2}^{t+\Delta t}f\sinh (\frac{3q_{2}}{2}\frac{^{t+\Delta
t}p}{^{t+\Delta t}\kappa })}
\end{equation}

\begin{equation}
\frac{\partial \Delta \lambda }{\partial ^{t+\Delta t}f}=\frac{^{t+\Delta
t}\kappa ^{ct}E^{v\mid e}}{3q_{1}q_{2}^{t+\Delta t}f^{2}\sinh (\frac{3q_{2}}{%
2}\frac{^{t+\Delta t}p}{^{t+\Delta t}\kappa })}
\end{equation}

\begin{equation}
\frac{\partial \Delta \lambda }{\partial ^{t+\Delta t}\kappa }=-\frac{%
^{ct}E^{v\mid e}}{3q_{1}q_{2}^{t+\Delta t}f\sinh (\frac{3q_{2}}{2}\frac{%
^{t+\Delta t}p}{^{t+\Delta t}\kappa })}-\frac{^{ct}E^{v\mid e}\cosh (\frac{%
3q_{2}}{2}\frac{^{t+\Delta t}p}{^{t+\Delta t}\kappa })^{t+\Delta t}p}{%
2q_{1}^{t+\Delta t}\kappa ^{t+\Delta t}f\sinh ^{2}(\frac{3q_{2}}{2}\frac{%
^{t+\Delta t}p}{^{t+\Delta t}\kappa })}
\end{equation}

\begin{equation}
\frac{\partial \Delta \lambda }{\partial ^{t+\Delta t}p}=\frac{^{ct}E^{v\mid
e}\cosh (\frac{3q_{2}}{2}\frac{^{t+\Delta t}p}{^{t+\Delta t}\kappa })}{%
2q_{1}^{t+\Delta t}f\sinh ^{2}(\frac{3q_{2}}{2}\frac{^{t+\Delta t}p}{%
^{t+\Delta t}\kappa })}
\end{equation}

If we separate the yield function (\ref{GTN function}) into two parts:

\begin{equation}
G_{l}:=\frac{\frac{3}{2}\left\Vert ^{t+\Delta t}\mathbf{T}^{d\mid
e}\right\Vert ^{2}}{^{t+\Delta t}\kappa ^{2}}
\end{equation}

\begin{equation}
G_{r}:=(1-q_{1}^{t+\Delta t}f)^{2}-2q_{1}^{t+\Delta t}f\text{ }[\cosh (\frac{%
3q_{2}}{2}\frac{^{t+\Delta t}p}{^{t+\Delta t}\kappa })-1]
\end{equation}

Because $G_{l}=Gr$, it should always stand that:

\begin{equation}
\frac{dG_{l}}{d_{0}^{t+\Delta t}\mathbf{E}_{e}}=\frac{dGr}{d_{0}^{t+\Delta t}%
\mathbf{E}_{e}}
\end{equation}

\begin{align}
&\frac{\partial G_{l}}{\partial ^{t+\Delta t}\mathbf{T}^{d\mid e}}\frac{%
d^{t+\Delta t}\mathbf{T}^{d\mid e}}{d^{t+\Delta t}\mathbf{T}_{e}}\frac{%
d^{t+\Delta t}\mathbf{T}_{e}}{d_{0}^{t+\Delta t}\mathbf{E}_{e}}+\frac{%
\partial G_{l}}{\partial ^{t+\Delta t}\kappa }\frac{d^{t+\Delta t}\kappa }{%
d\Delta \lambda }\frac{d\Delta \lambda }{d_{0}^{t+\Delta t}\mathbf{E}_{e}} 
\notag \\
&=\frac{\partial G_{r}}{\partial ^{t+\Delta t}\kappa }\frac{d^{t+\Delta
t}\kappa }{d\Delta \lambda }\frac{d\Delta \lambda }{d_{0}^{t+\Delta t}%
\mathbf{E}_{e}}+\frac{\partial G_{r}}{\partial ^{t+\Delta t}p}\frac{%
d^{t+\Delta t}p}{d^{t+\Delta t}\mathbf{T}_{e}}\frac{d^{t+\Delta t}\mathbf{T}%
_{e}}{d_{0}^{t+\Delta t}\mathbf{E}_{e}}+\frac{\partial G_{r}}{\partial
^{t+\Delta t}f}\frac{d^{t+\Delta t}f}{d_{0}^{t+\Delta t}\mathbf{E}_{e}}
\label{(2)}
\end{align}

After rewriting (\ref{(2)}), we can get:

\begin{align}
\frac{\partial G_{r}}{\partial ^{t+\Delta t}f}\frac{d^{t+\Delta t}f}{%
d_{0}^{t+\Delta t}\mathbf{E}_{e}}& =(\frac{\partial G_{l}}{\partial
^{t+\Delta t}\kappa }\frac{d^{t+\Delta t}\kappa }{d\Delta \lambda }-\frac{%
\partial G_{r}}{\partial ^{t+\Delta t}\kappa }\frac{d^{t+\Delta t}\kappa }{%
d\Delta \lambda })\frac{d\Delta \lambda }{d_{0}^{t+\Delta t}\mathbf{E}_{e}} 
\notag \\
& +\frac{\partial G_{l}}{\partial ^{t+\Delta t}\mathbf{T}^{d\mid e}}\frac{%
d^{t+\Delta t}\mathbf{T}^{d\mid e}}{d^{t+\Delta t}\mathbf{T}_{e}}\frac{%
d^{t+\Delta t}\mathbf{T}_{e}}{d_{0}^{t+\Delta t}\mathbf{E}_{e}}-\frac{%
\partial G_{r}}{\partial ^{t+\Delta t}p}\frac{d^{t+\Delta t}p}{d^{t+\Delta t}%
\mathbf{T}_{e}}\frac{d^{t+\Delta t}\mathbf{T}_{e}}{d_{0}^{t+\Delta t}\mathbf{%
E}_{e}}
\end{align}

The related derivatives are \ref{dk_ddlam} and:

\begin{equation}
\frac{\partial G_{l}}{\partial ^{t+\Delta t}\mathbf{T}^{d\mid e}}=\frac{%
3^{t+\Delta t}\mathbf{T}^{d\mid e}}{^{t+\Delta t}\kappa ^{2}}
\end{equation}

\begin{equation}
\frac{\partial G_{l}}{\partial ^{t+\Delta t}\kappa }=-\frac{3\left\Vert
^{t+\Delta t}\mathbf{T}^{d\mid e}\right\Vert ^{2}}{^{t+\Delta t}\kappa ^{3}}
\end{equation}

\begin{equation}
\frac{\partial G_{r}}{\partial ^{t+\Delta t}\kappa }=\frac{%
q_{1}q_{2}^{t+\Delta t}f^{t+\Delta t}p\sinh (\frac{3q_{2}}{2}\frac{%
^{t+\Delta t}p}{^{t+\Delta t}\kappa })}{^{t+\Delta t}\kappa ^{2}}
\end{equation}

\begin{equation}
\frac{\partial G_{r}}{\partial ^{t+\Delta t}f}=-2q_{1}\cosh (\frac{3q_{2}}{2}%
\frac{^{t+\Delta t}p}{^{t+\Delta t}\kappa })
\end{equation}

\begin{equation}
\frac{\partial G_{r}}{\partial ^{t+\Delta t}p}=-\frac{3q_{1}q_{2}^{t+\Delta
t}f\sinh (\frac{3q_{2}}{2}\frac{^{t+\Delta t}p}{^{t+\Delta t}\kappa })}{%
^{t+\Delta t}\kappa }
\end{equation}

From (\ref{(1)}) we can get:

\begin{align}
\frac{d^{t+\Delta t}f}{d_{0}^{t+\Delta t}\mathbf{E}_{e}}&=(\frac{\partial
G_{r}}{\partial ^{t+\Delta t}f})^{-1}(\frac{\partial G_{l}}{\partial
^{t+\Delta t}\kappa }-\frac{\partial G_{r}}{\partial ^{t+\Delta t}\kappa })%
\frac{d^{t+\Delta t}\kappa }{d\Delta \lambda }\frac{d\Delta \lambda }{%
d_{0}^{t+\Delta t}\mathbf{E}_{e}}  \notag \\
&+(\frac{\partial G_{r}}{\partial ^{t+\Delta t}f})^{-1}(\frac{\partial G_{l}%
}{\partial ^{t+\Delta t}\mathbf{T}^{d\mid e}}\frac{d^{t+\Delta t}\mathbf{T}%
^{d\mid e}}{d^{t+\Delta t}\mathbf{T}^{\mid e}}-\frac{\partial G_{r}}{%
\partial ^{t+\Delta t}p}\frac{d^{t+\Delta t}p}{d^{t+\Delta t}\mathbf{T}%
^{\mid e}})\frac{d^{t+\Delta t}\mathbf{T}^{\mid e}}{d_{0}^{t+\Delta t}%
\mathbf{E}_{e}}  \label{(3)}
\end{align}

After substituting the expression of $\frac{d^{t+\Delta t}f}{d_{0}^{t+\Delta
t}\mathbf{E}_{e}}$ of (\ref{(3)}) into (\ref{(2)}), we can get the
expression for$\frac{d\Delta \lambda }{d_{0}^{t+\Delta t}\mathbf{E}_{e}}$%
from:

\begin{align}
& {\small [1-}\frac{\partial \Delta \lambda }{\partial ^{t+\Delta t}\kappa }%
\frac{d^{t+\Delta t}\kappa }{d\Delta \lambda }{\small -(}\frac{\partial
\Delta \lambda }{\partial ^{ct}E^{v\mid e}}\frac{d^{ct}E^{v\mid e}}{%
d^{t+\Delta t}f}{\small +}\frac{\partial \Delta \lambda }{\partial
^{t+\Delta t}f}{\small )(}\frac{\partial G_{r}}{\partial ^{t+\Delta t}f}%
{\small )}^{-1}{\small (}\frac{\partial G_{l}}{\partial ^{t+\Delta t}\kappa }%
{\small -}\frac{\partial G_{r}}{\partial ^{t+\Delta t}\kappa }{\small )}%
\frac{d^{t+\Delta t}\kappa }{d\Delta \lambda }{\small ]}\frac{d\Delta
\lambda }{d_{0}^{t+\Delta t}\mathbf{E}_{e}}  \notag \\
& =(\frac{\partial \Delta \lambda }{\partial ^{ct}E^{v\mid e}}\frac{%
d^{ct}E^{v\mid e}}{d^{t+\Delta t}f}+\frac{\partial \Delta \lambda }{\partial
^{t+\Delta t}f})(\frac{\partial G_{r}}{\partial ^{t+\Delta t}f})^{-1}(\frac{%
\partial G_{l}}{\partial ^{t+\Delta t}\mathbf{T}^{d\mid e}}\frac{d^{t+\Delta
t}\mathbf{T}^{d\mid e}}{d^{t+\Delta t}\mathbf{T}^{\mid e}}-\frac{\partial
G_{r}}{\partial ^{t+\Delta t}p}\frac{d^{t+\Delta t}p}{d^{t+\Delta t}\mathbf{T%
}^{\mid e}})\frac{d^{t+\Delta t}\mathbf{T}^{\mid e}}{d_{0}^{t+\Delta t}%
\mathbf{E}_{e}}  \notag \\
& +\frac{\partial \Delta \lambda }{\partial ^{t+\Delta t}p}\frac{d^{t+\Delta
t}p}{d^{t+\Delta t}\mathbf{T}^{\mid e}}\frac{d^{t+\Delta t}\mathbf{T}^{\mid
e}}{d_{0}^{t+\Delta t}\mathbf{E}_{e}}
\end{align}

After getting $\frac{d\Delta \lambda }{d_{0}^{t+\Delta t}\mathbf{E}_{e}}$, $%
\frac{d^{t+\Delta t}f}{d_{0}^{t+\Delta t}\mathbf{E}_{e}}$ can be obtained
from (\ref{(3)}). Once $\frac{d\Delta \lambda }{d_{0}^{t+\Delta t}\mathbf{E}%
_{e}}$and $\frac{d^{t+\Delta t}f}{d_{0}^{t+\Delta t}\mathbf{E}_{e}}$are
known, $\frac{d^{ct}\mathbf{E}^{d\mid e}}{d\,_{0}^{t+\Delta t}\mathbf{E}_{e}}
$can be obtained from (\ref{dEectdev_dEe}), thus $\frac{d\,^{ct}\mathbf{E}%
_{e}}{d\,_{0}^{t+\Delta t}\mathbf{E}_{e}}$can be obtained from (\ref%
{dEect_dEe}), and $\frac{d\,_{0}^{t+\Delta t}\mathbf{E}_{e}}{d\,^{tr}\mathbf{%
E}_{e}}$can be obtained from (\ref{dEe_dEetr}). When $\frac{%
d\,_{0}^{t+\Delta t}\mathbf{E}_{e}}{d\,^{tr}\mathbf{E}_{e}}$ is known, $%
^{t+\Delta t}\mathbb{A}_{ep}^{|tr}=\frac{d\,^{t+\Delta t}\mathbf{T}^{|tr}}{%
d\,^{tr}\mathbf{E}_{e}}\simeq \frac{d\,^{t+\Delta t}\mathbf{T}^{|e}}{d\,^{tr}%
\mathbf{E}_{e}}=\frac{d\,^{t+\Delta t}\mathbf{T}^{|e}}{d\,_{\hspace{3.2ex}%
0}^{t+\Delta t}\mathbf{E}_{e}}:\frac{d\,_{\hspace{3.2ex}0}^{t+\Delta t}%
\mathbf{E}_{e}}{d\,^{tr}\mathbf{E}_{e}}=\,^{t+\Delta t}\mathbb{A}^{|e}:\frac{%
d\,_{\hspace{3.2ex}0}^{t+\Delta t}\mathbf{E}_{e}}{d\,^{tr}\mathbf{E}_{e}}$
is known.

\subsection{Numerical simulations}

GTN model with current framework has been implemented as an user-material
subroutine in our in-house finite element code DULCINEA. In the following
numerical examples, we used single-element simulations to demonstrate the
influence of the parameters $q_{1}$ and $q_{2}$, and to discuss if the
elastic constants of the continuum should change due to the change of the
volume fraction of voids. We simulated the tensile test of a cube to show
the influence of the hardening parameter $H$, as well as to show the
robustness of the algorithm. Furthermore, we ran the tensile test of a
necking bar, and prescribed the von Mises stress contours plot, the void
volume fraction contour plot as well as the displacement-reaction force
curve. The results are reasonable and the algorithm is robust.

\subsubsection{Tensile test of a single element}

Because the main purpose of this work is to introduce a new framework for
the GTN\ model, instead of improving the model by adding extra rules for
accounting for phenomena like void nucleation or void coalescence, only the
GTN yield function is used as the constitutive rule. The parameters are only 
$q_{1}$ and $q_{2}$. Following the traditional values given to $q_{1}$ and $%
q_{2}$, the following is a brief study of the influence of the two
parameters. Here no isotropic hardening is included, and the material
parameters are listed in table\ \ref{table_elastic_constants}.

\begin{table}[tbp]
\centering%
\begin{tabular}{|c|c|c|c|}
\hline
$C_{11}$ & $C_{12}$ & $C_{44}$ & $k_{0}$ \\ \hline
$269.4$ GPa & $115.4$ GPa & $77.0$ GPa \  & $96.0$ MPa \\ \hline
\end{tabular}%
\caption{The elastic constants and the initial yield stress of the matrix
material. }
\label{table_elastic_constants}
\end{table}

\begin{figure}[tbp]
\centering
\includegraphics[width=1.0\textwidth]{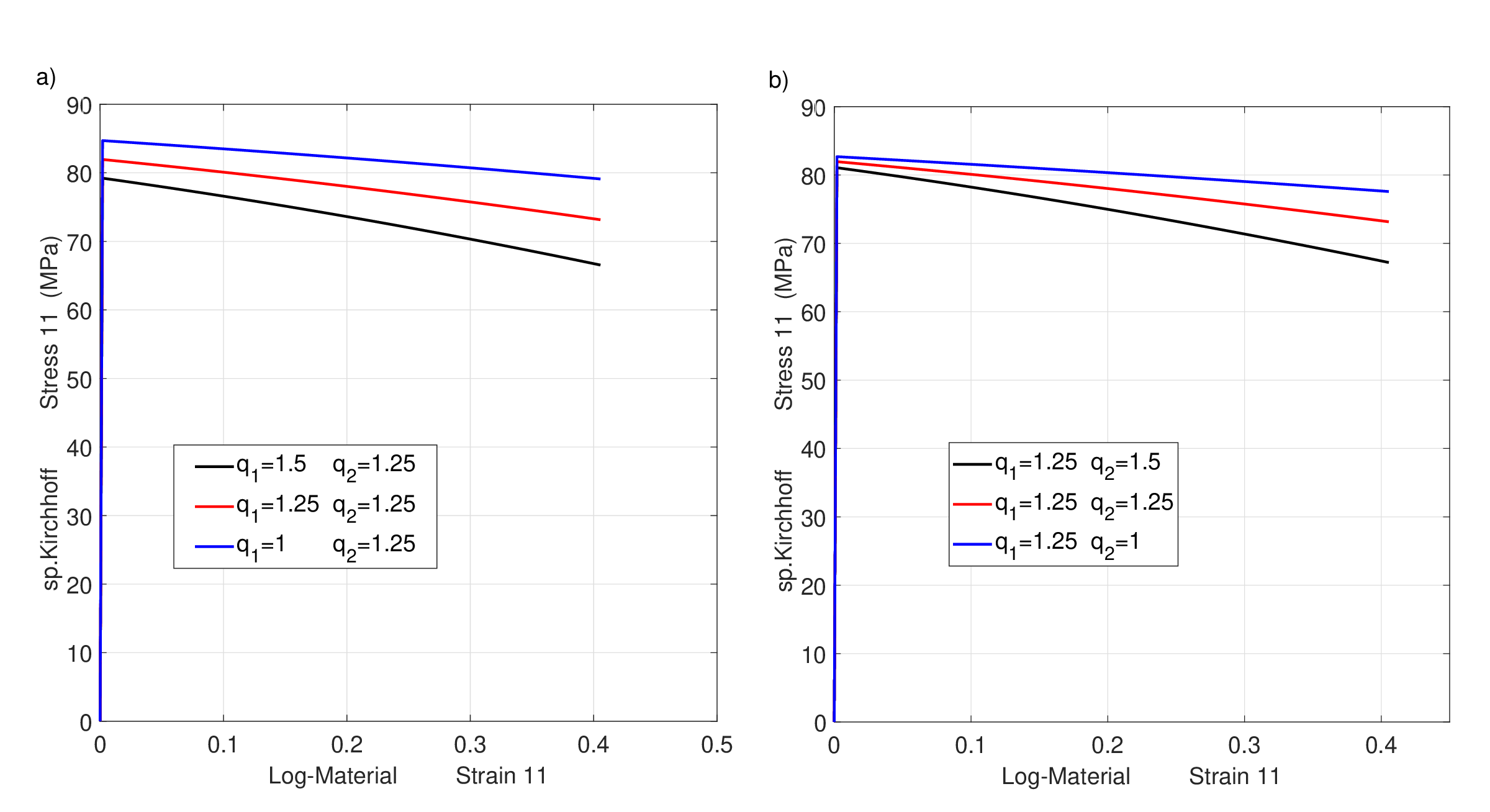}
\caption{a) the influence of parameter $q_{1}$; b) the influence of
parameter $q_{2}$}
\label{q1_q2}
\end{figure}

Similar to previous GTN models, by increasing $q_{1}$, the void growth would
be accelerated, and the flow stress would be decreased. By increasing $q_{2}$%
, the void growth rate would also be increased. When isotropic hardening is
not included, the flow stress reflects the void volume fraction, so the
influence of $q_{1}$ and $q_{2}$ is shown in figure \ref{q1_q2}. The
simulation is a uniaxial tensile test with one element, with homogeneous
boundary conditions with only the loading direction fully constraint. The
initial volume fraction of void $f_{0}$ is set as $0.1$.

\begin{figure}[tbp]
\centering
\includegraphics[width=1.0\textwidth]{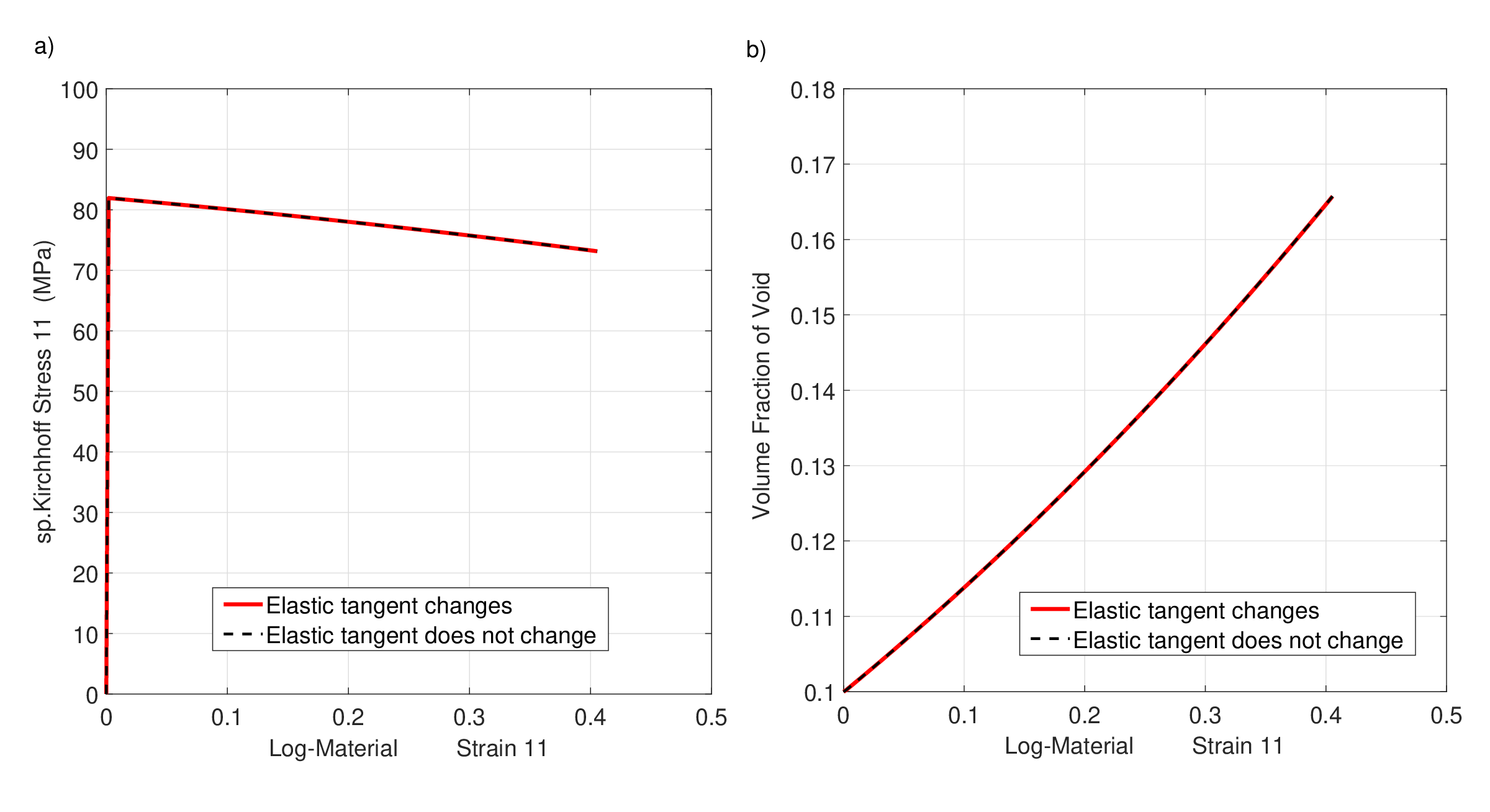}
\caption{a) the calculated flow stress and b) the calculated volume fraction
in the cases that the elastic tangent of the continuum changes and does not
change according to the volume fraction of the voids}
\label{elastic_tangent_plot}
\end{figure}

It should be noticed that because the elastic tangent of the void should be
zero, the elastic tangent of the continuum should change according to the
change of the volume fraction of the void. It is noticed that with current
GTN model, this change does not influence the calculated flow stress and the
volume fraction of the void, as shown in figure \ref{elastic_tangent_plot},
however, the elastic strain as well as the stored energy are different in
the two cases, so treating the elastic tangent of the continuum as constant
in GTN models is problematic even though it does not change the results of
flow stress or volume fraction of voids. For the simulation in the plot \ref%
{elastic_tangent_plot}, $q_{1}=1.25$, $q_{2}=1.25$.

\subsubsection{Tensile test of a cube}

The sample is cubic with size length 1mm. Prescribed tensile displacement up
to $0.5$ mm is applied on the $x$ direction, and homogeneous boundary
condition, which allows the sample to change freely on the other two
directions. The undeformed and deformed sample are shown in figure \ref%
{cube_shape}. The element used is C3D20R, which has $20$ nodes and $8$
integration points. The total element number is $512$. The total loading
steps is $250$. The material parameters used are listed in the table \ref%
{table_elastic_constants}, with initial void volume fraction $f_{0}=0.05$
and parameters for the model $q_{1}=1$, $q_{2}=1.25$.

For this test, we set the isotropic hardening parameter $H$ to different
values, and compared the averaged stress-strain curves (a) of figure \ref%
{cube_H} as well as the void volume fraction-strain curves (b) of figure \ref%
{cube_H}, from which we can conclude that for adding the isotropic
hardening, when the yield stress of matrix $\kappa $ is related to $\lambda $%
, it would not influence the void volume fraction.

In the table \ref{table_convergence_cube}, we show the global convergence
conditions of the force and energy of several typical steps for the case
when no isotropic hardening is included.

\begin{figure}[tbp]
\centering
\includegraphics[width=1.0\textwidth]{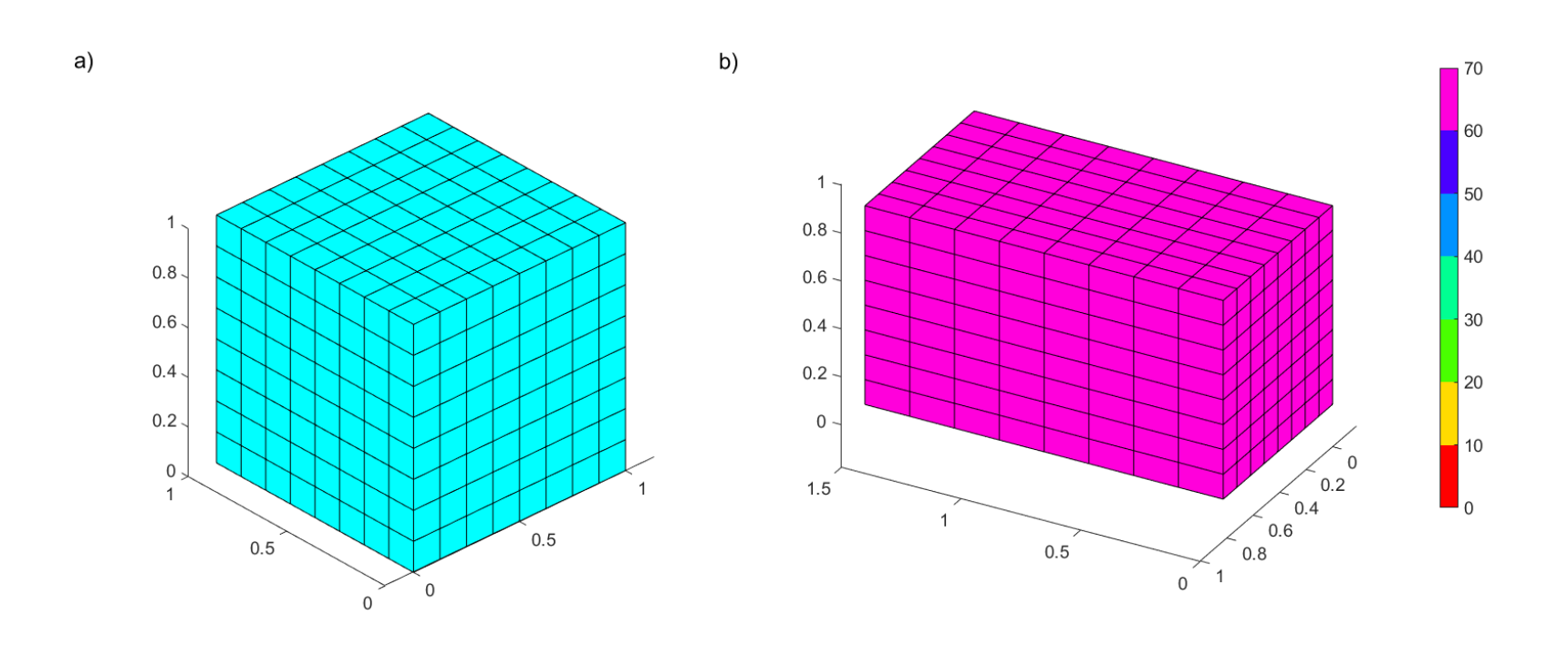}
\caption{a) the undeformed shape of the cubic; b) the von Mises stress state
of the deformed cubic.}
\label{cube_shape}
\end{figure}

\begin{figure}[tbp]
\centering
\includegraphics[width=1.0\textwidth]{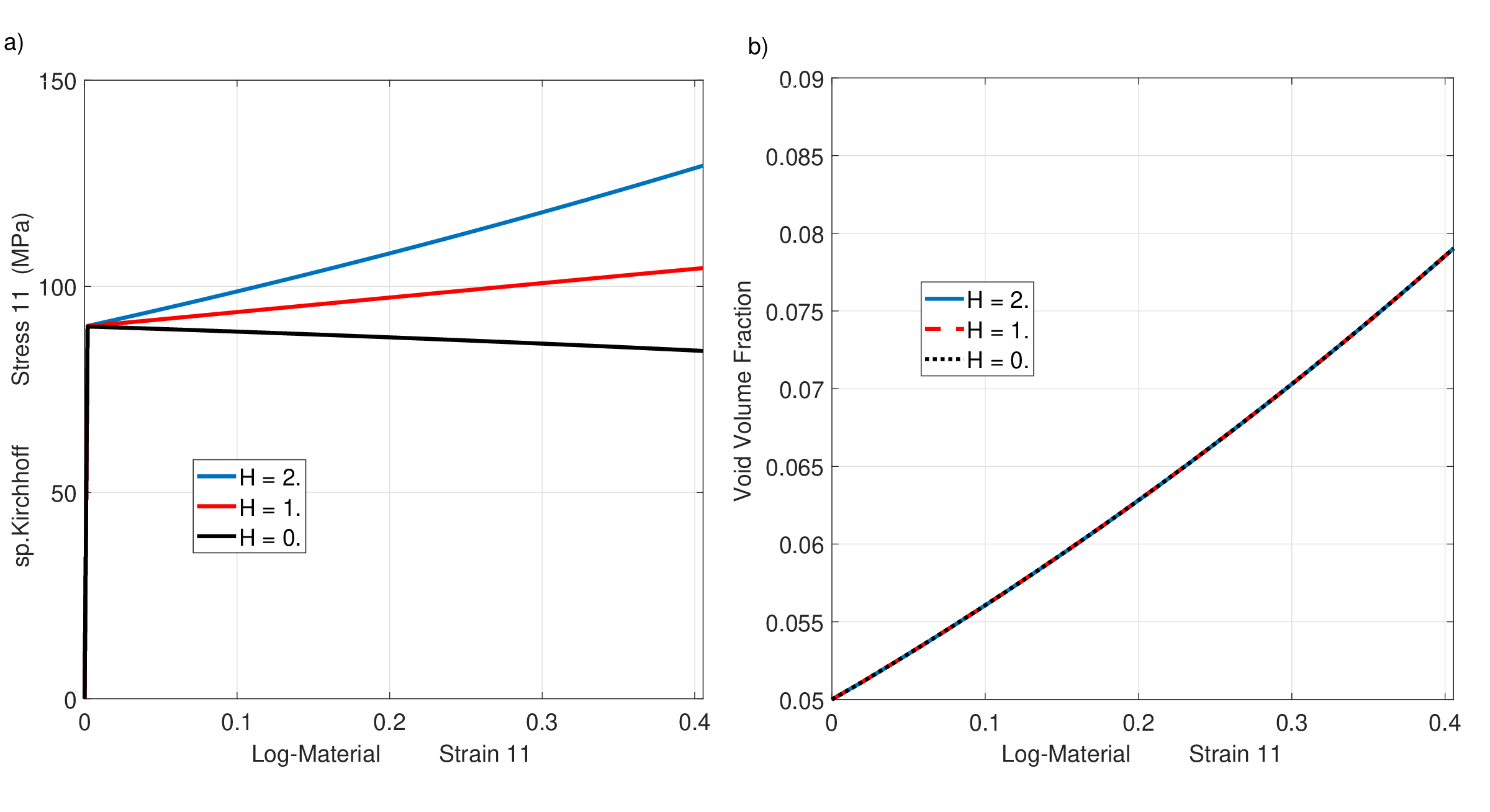}
\caption{a) the averaged stress-strain curves for the cases when the
hardening parameter $H$ has different values; b) the averaged void volume
fraction-strain curves for the cases when the hardening parameter $H$ has
different values}
\label{cube_H}
\end{figure}

\begin{table}[tbp]
{\footnotesize \centering%
\begin{tabular}{ccccccc}
& \multicolumn{3}{c}{Residual norm} & \multicolumn{3}{c}{Energy norm} \\ 
Iteration & \multicolumn{3}{c}{Step} & \multicolumn{3}{c}{Step} \\ 
& 75 & 150 & 225 & 75 & 150 & 225 \\ 
$\left( 1\right) $ & $5.417E+02$ & $5.486E+02$ & $5.649E+02$ & $1.566E+01$ & 
$1.587E+01$ & $1.635E+01$ \\ 
$\left( 2\right) $ & $4.517E-02$ & $3.583E-02$ & $2.937E-02$ & $5.069E-07$ & 
$3.091E-07$ & $1.988E-07$ \\ 
$\left( 3\right) $ & $5.085E-05$ & $3.411E-05$ & $2.404E-05$ & $2.060E-12$ & 
$3.105E-12$ & $6.069E-12$%
\end{tabular}%
}
\caption{Global convergence for several typical steps.}
\label{table_convergence_cube}
\end{table}

\subsubsection{Tensile test of a plate with a hole}

\begin{figure}[tbp]
\centering
\includegraphics[width=1.0\textwidth]{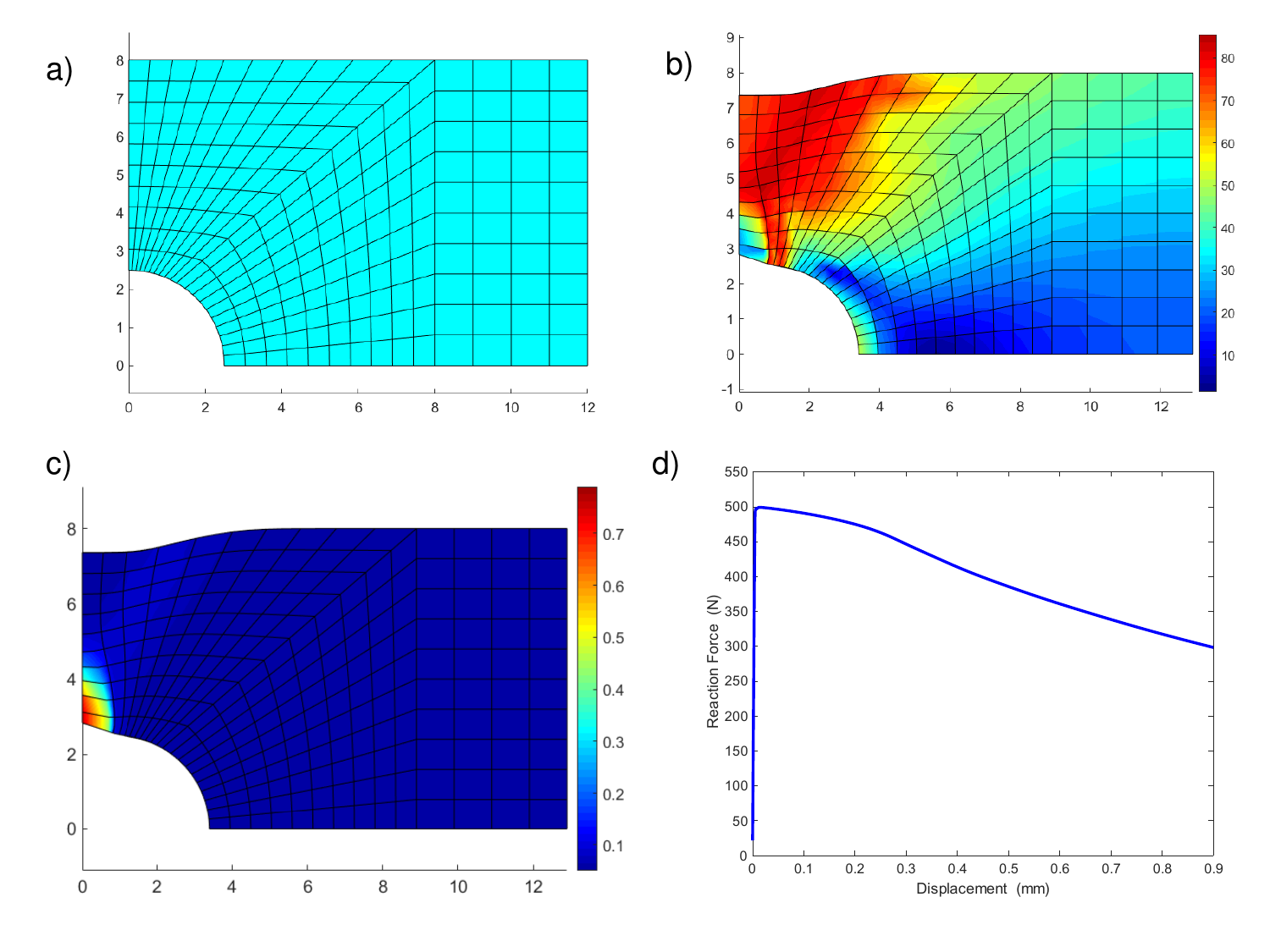}
\caption{a) the dimensions of the undeformed plate; b) the deformed plate
with contour plot of von Mises stress; c) the deformed plate with contour
plot of the volume fraction of void; d) the displacement-reaction force plot
of the loading end}
\label{plate_4pics}
\end{figure}

To further test the model, we ran the classic example of the tensile test of
a thin plate with a hole in the middle. The thickness of the plate is 1 mm,
and the other dimensions are shown in its undeformed shape in figure a) of %
\ref{plate_4pics}. The element C3D20R is used, because of symmetry, only $%
1/4 $ of the plate is modelled. To avoid early localization of deformation,
a linear isotropic hardening with $H=0.85$ is added: $\kappa =\kappa
_{0}+0.85\lambda $. For the model, the other material parameters used are
also from the table \ref{table_elastic_constants}, with initial void volume
fraction $f_{0}=0.05$ and parameters for the model $q_{1}=1$, $q_{2}=1.25$.

The total tensile displacement is $7.5\%$ of the length of the plate. The
von Mises stress condition as well as the distribution of volume fraction of
void are plotted in figure b) and c) of \ref{plate_4pics}. The reaction
force-displacement curve of the loading end of the specimen is plotted in d)
of \ref{plate_4pics}. For this work, we do not consider the coalescence of
voids or the fracture criteria.

\section{Conclusion}

\bigskip In this work, we demonstrate a new large-strain formulation for
void evolution mechanisms. This formulation is an extension of our
established framework that uses elastic correctors and logarithmic strains.
It adopts the advantages of the framework, for example, it is fully
hyperelastic and maintains the Kroner-Lee multiplicative decomposition;
there is no constraint on the amount of the elastic strain and the form of
the stored strain energy; both the theory and algorithm have an additive
structure; it is consistent and parallel to continuum elastoplasticity as
well as crystal plasticity. We proved that from the kinematics, we can get
an accurate function$~$that relates the evolution of the volume fraction of
voids and the evolution of the corrector of the elastic volumetric strain $%
^{ct}\dot{E}_{e}^{v}=-\frac{\dot{f}}{1-f}$. We use the GTN yield function as
an example of the void evolution rule to demonstrate the implementation of
this formulation. With some numerical examples, we show that the implemented
model can capture the features of the GTN model, and the algorithm is robust
and simple.

\bibliographystyle{model2-names}
\bibliography{literature}

\end{document}